%% file: psync.tex
\documentclass[10pt]{article}

\usepackage{fullpage}
\usepackage{amssymb,amsmath}
\usepackage{amsfonts}
\usepackage{epsfig}
\usepackage{color}
\usepackage{algorithm}
\usepackage[noend]{algpseudocode}
\usepackage{subfigure}
\usepackage{def}

\input amssym.def

\title{\bf Partially Ordered Distributed Computations on Asynchronous Point-to-Point Networks\thanks{This work is partially supported by CNPq/PADCT.}}
\author{{\sc Ricardo C. Corr\^ea}\thanks{Partially supported by CNPq. This work was done when this author was visiting the Programa de Engenharia de Sistemas e Computa\c c\~ao, COPPE, Universidade Federal do Rio de Janeiro.} \\
Universidade Federal do Cear\'a \\
Departamento de Computa\c c\~ao \\
Campus do Pici, Bloco 910,
60455-760 Fortaleza - CE, Brazil \\
{\tt correa@lia.ufc.br} \and {\sc Valmir C. Barbosa}\thanks{Partially supported by CNPq, CAPES, and a FAPERJ BBP grant.} \\
Universidade Federal do Rio de Janeiro \\
Programa de Engenharia de Sistemas e Computa\c c\~ao, COPPE\\
Caixa Postal 68511,
21941-972 Rio de Janeiro - RJ, Brazil\\
{\tt valmir@cos.ufrj.br}}

\date{}

\begin{document} 

\maketitle

\begin{abstract} 
Asynchronous executions of a distributed algorithm differ from each other due to the nondeterminism in the order in which the messages exchanged are handled. In many situations of interest, the asynchronous executions induced by restricting nondeterminism are more efficient, in an application-specific sense, than the others. In this work, we define partially ordered executions of a distributed algorithm as the executions satisfying some restricted orders of their actions in two different frameworks, those of the so-called event- and pulse-driven computations. The aim of these restrictions is to characterize asynchronous executions that are likely to be more efficient for some important classes of applications. Also, an asynchronous algorithm that ensures the occurrence of partially ordered executions is given for each case. Two of the applications that we believe may benefit from the restricted nondeterminism are backtrack search, in the event-driven case, and iterative algorithms for systems of linear equations, in the pulse-driven case.

\end{abstract} 

\input{intro}

\input{application}
\input{event}
\input{pulse}
\input{conc}

\bibliography{/home/correa/artigos/bib/distributed}
\bibliographystyle{plain}

\end{document}

%% file: intro.tex
\section{Introduction}
\label{sec:intro}

We consider a system represented by a connected undirected graph $G = (N, E)$, where $N = \{ 1, 2, \cdots, n \}$ stands for a set of nodes and $E$ for a set of point-to-point bidirectional communication channels. A channel involving two distinct nodes $i$ and $j$ is denoted by $ij$. For $i \in N$, we let $N(i) = \{ j \mid ij \in E \}$ comprise the {\em neighbors} of $i$ in $G$. In the system represented by $G$, a node is able to sequentially perform computations and to interact with neighbors solely by sending or receiving messages through the channels incident to it. Every node has its own local, independent clock, but has no access to a global clock of any kind. All channels are reliable, which means that every message is delivered to its destination with finite but unpredictable delay. This configuration characterizes the distributed and asynchronous nature of the system represented by $G$.

A distributed computation carried out on $G$ is fully described by an initial global state (comprising an initial state for each node and no messages in transit), the local computations performed by each node, and the interactions among nodes. The local computation of a node and the messages received from neighbors in $G$ determine the evolution of its local state. Motivated by a number of applications, we consider two categories of distributed computations, depending upon what governs the local state transitions of the nodes. In the first category, that of {\em event-driven computations}, each node reacts whenever it receives a message by performing a local computation, as shown in Algorithm~\ref{alg:event-driven}. More precisely, the receiving of a message from a neighbor  affects node $i$'s local state by means of the execution of procedure \Call{event$_i$}{}, which encapsulates the actions of the particular computation associated with node $i$. Besides changing the local state of $i$, the execution of \Call{event$_i$}{} produces as a result a set $MSG_i$ of messages, possibly empty. Each message in $MSG_i$, if any, has one of $i$'s neighbors as destination. This framework is widely adopted in the description and analysis of asynchronous distributed algorithms for several applications~\cite{Barbosa.96}.

\begin{algorithm}[htbp]
\caption{Outline of the computation at node $i$ in the event-driven framework.}
\label{alg:event-driven}
\begin{algorithmic}[1]
\State Set initial state of $i$ \label{lin:controlvar}
\State \Call{event$_i$}{ - , $MSG_i$}
\State Send each message in $MSG_i$ to a specified neighbor \label{lin:onesend} 
\While {global termination is not known to $i$}
        \If {$msg_i$ has arrived from a neighbor of $i$} \label{lin:msgdeliveryevent}
		\State \Call{event$_i$}{$msg_i, MSG_i$}
		\State Send each message in $MSG_i$ to a specified neighbor \label{lin:anothersend}
	\EndIf
\EndWhile
\end{algorithmic}
\end{algorithm}

The second category is that of {\em pulse-driven computations}, and is motivated by applications in which the initial state evolves towards a final state in phases, as in many iterative algorithms. To model such a behavior, we assume that a mechanism that generates a sequence of pulses is provided to govern evolution at each node $i$. Such a mechanism is given in Algorithm~\ref{alg:pulse-driven} by two functions, namely \Call{getCurrent$_i$}{} and \Call{hasAdvanced$_i$}{}. The former is used to notify the pulse generation mechanism that node $i$ has started the local computations associated with the most recent pulse generated. On the other hand, the pulse generation mechanism signals the generation of a new pulse with the Boolean function \Call{hasAdvanced$_i$}{}. An additional assumption is that if \Call{hasAdvanced$_i$}{} returns $\mathbf{true}$, then it does not return $\mathbf{true}$ again before \Call{getCurrent$_i$}{} is invoked. The role of procedure \Call{event$_i$}{} is only to incorporate any relevant information contained in $msg_i$, whereas the local state transition is performed by procedure \Call{pulse$_i$}{} (which also gets the current pulse as part of its input).

\begin{algorithm}[htbp]
\caption{Outline of the computation at node $i$ in the pulse-driven framework.}
\label{alg:pulse-driven}
\begin{algorithmic}[1]
\State Set the initial state of $i$ 
\State $\ell_i \gets$ \Call{getCurrent$_i$}{\ }
\State \Call{pulse$_i$}{$\ell_i$, $MSG_i$}
\State Send each message in $MSG_i$ to a specified neighbor \label{lin:onesend2}
\While {global termination is not known to $i$}
	\If {\Call{hasAdvanced$_i$}{\ }} \label{lin:hasadvanced}
		\State $\ell_i \gets$ \Call{getCurrent$_i$}{\ }
		\State \Call{pulse$_i$}{$\ell_i$, $MSG_i$}
		\State Send each message in $MSG_i$ to a specified neighbor \label{lin:anothersend2}
        \ElsIf {$msg_i$ has arrived from a neighbor of $i$} \label{lin:msgdeliverypulse}
		\State \Call{event$_i$}{$msg_i$}
	\EndIf
\EndWhile
\end{algorithmic}
\end{algorithm}

Distributed computations described by the frameworks above are nondeterministic for two reasons, namely the asynchronous local computations on the various nodes and the nondeterministic delays incurred by message transfers through the network. These two factors affect the order in which messages are processed by their destination nodes and, consequently, the evolution of the nodes' local states. In many situations of interest, computations starting at a fixed initial state evolve more rapidly towards a final state than others (the actual meaning of ``more rapidly'' may be faster or more efficient according to some performance criterion other than time). Thus, it is sometimes desirable to restrict the nondeterminism of a distributed computation, which can be accomplished by an appropriate control of the message ordering or the pulse generation mechanisms. For instance, synchronous distributed branch-and-bound may visit fewer subproblems than the asynchronous version~\cite{Correa.Ferreira.95,Karp.Zhang.93,Li.Wah.86}; in several cases of distributed iterative algorithms for solving systems of equations, convergence is not guaranteed for asynchronous executions, unless some restrictions on the order of the messages are respected~\cite{Bertsekas.Tsitsiklis.89}; other examples range from multimedia to agent systems~\cite{Adelstein.Singhal.97,Dobrev.Flocchini.Prencipe.Santoro.06,Dobrev.Kralovic.Santoro.Shi.06}. 

Motivated by the fact that reducing the nondeterminism of distributed computations is useful in a wide variety of applications, restricted message ordering and synchronization mechanisms have been implemented in several systems~\cite{Birman.Joseph.87,Birman.Schiper.Stephenson.91,Peterson.Bucholz.Schlichting.89}. These mechanisms implement two types of condition. The first type is associated with the order in which messages are delivered to their destinations~\cite{Kshemkalyani.Singhal.98}. In general terms, such a condition may be stated as follows:
\begin{description}
\item[Message Delivery Condition:] No message $msg$ is to be accepted by node $i \in N$ until all local actions that are required to occur before the reception of $msg$ have taken place. Reception of $msg$ by $i$ may then have to be postponed.
\end{description}
The canonical example here is the FIFO ordering, which requires that every message be delivered to its destination only after all other messages sent before it by the same sender through the same channel. The second type of condition is specific to pulse-driven computations and targets the control of nondeterminism via the pulse generation mechanism:
\begin{description}
\item[Pulse Generation Condition:] No pulse is to be generated at node $i \in N$ until all local actions that are required to occur up to the current pulse have taken place. Generation of a new pulse at $i$ may then have to be postponed.
\end{description}
Particular cases of this condition are present in many models of synchronous distributed computation or in algorithms for simulating such models~\cite{Barbosa.96,Lynch.96}.

In this paper, we deal with partially ordered computations constituting the subset of all possible asynchronous distributed computations that comprise only those satisfying some restricted order of their actions. In the event-driven framework, these restricting orders affect the order in which messages are delivered to their destinations (at line~\ref{lin:msgdeliveryevent} of Algorithm~\ref{alg:event-driven}). In this context, we give a new message delivery condition that generalizes the one leading to the FIFO ordering in the sense that the set of asynchronous computations satisfying the new condition may be larger than the original one. The computations that violate the FIFO ordering but are admitted by the new condition depend on parameters dynamically adjustable during these computations. In the same sense, we also generalize the condition associated with the causal ordering studied in~\cite{Birman.Joseph.87,Garg.04}. The principle of defining an ordering that can be dynamically tuned to become more or less strict as the computation evolves can also be applied to the pulse-driven framework. In this case, a generalization of an ordering can be obtained when the relation between the pulse generation mechanism and message delivery (lines~\ref{lin:hasadvanced} and~\ref{lin:msgdeliverypulse} of Algorithm~\ref{alg:pulse-driven}) is required to satisfy some constraints. We introduce an ordering that generalizes the well-known fully synchronous ordering~\cite{Barbosa.96}, among others.

Besides this introductory section, we present in Section~\ref{sec:application} motivating applications for later discussion. Two general distributed problem solving methods are used for this purpose. The usefulness of a generalized message delivery condition, and of the ordering it induces, for the event-driven framework is illustrated by an asynchronous version of the synchronous distributed randomized backtrack search algorithm of~\cite{Karp.Zhang.93} in which the asynchronism is controlled. For the pulse-driven framework, implementations of iterative algorithms for systems of linear equations are also presented in that section. The generalized message delivery conditions we propose for the event-driven framework are then presented in Section~\ref{sec:event}, whereas the pulse-driven computations, with the new message delivery and pulse generation conditions, are the subject of Section~\ref{sec:pulse}. In both sections, the use of the new conditions in the motivating applications is also discussed. We close the paper with some concluding remarks and directions for further work in Section~\ref{sec:conc}.

%% file: application.tex
\section{Two motivating applications}
\label{sec:application}

In this section, we describe applications in which the local state of each node $i \in N$ evolves according to the computations performed locally by $i$, which in turn are affected by the local states of other nodes. For each application, we give a brief description using the frameworks mentioned in Section~\ref{sec:intro}, and we outline some partially ordered executions that have the effect of establishing dependencies between local states in distinct nodes in terms of the message delivery and pulse generation conditions. Further details of their implementations are left to Sections~\ref{sec:event} and~\ref{sec:pulse}.

\subsection{Backtrack search}
\label{sec:backtrack}

\subsubsection{Search problems}

A {\em search problem} asks that certain object arrangements, called {\em solutions}, be found out of a large set of arrangements. Enumerating the extensions of a partially ordered set is a typical search problem, in which the arrangements are all the binary relations that can be defined on a set given as input and the solutions are only the relations that are partial orders containing the given partial order on the input set~\cite{Correa.Szwarcfiter.05}. Search problems are amenable to a distributed treatment when the set of possible arrangements can be appropriately partitioned among the nodes of the distributed system. We take as our first example the randomized approach proposed and analyzed in~\cite{Karp.Zhang.93} for such a partitioning in backtrack search algorithms. In the context of the algorithm in~\cite{Karp.Zhang.93}, henceforth referred to as the Karp-Zhang algorithm, a backtrack algorithm proceeds by successively applying a {\em branching procedure} to partition the set of possible arrangements. During this process, a {\em subproblem} corresponds to a subset of the set of possible arrangements. Each time the branching procedure is applied to subproblem $s$, it either solves $s$ directly (and does not produce any other subproblem) or derives from $s$ a set of subproblems such that the solutions of $s$ can be found from the solutions of the derived subproblems. An assumption that is tacitly made is that the branching procedure produces a tree, defined by the subproblems as nodes and the relation ``subproblem derived from a branching of'' as edges. We consider the case in which {\em all} leaves of the search tree must be generated and solved by the branching procedure. Thus, the recursion stops only when there are no subproblems left.

\subsubsection{An event-driven randomized algorithm}

The Karp-Zhang algorithm is randomized and may be described in the event-driven framework as follows. At any point of the execution, a {\em frontier subproblem} is a subproblem that has been generated but not yet passed on to the branching procedure. The intrinsic concurrency of the backtrack search stems from the fact that frontier subproblems can be distributed among the nodes, each subproblem to exactly one node. The frontier subproblems assigned to node $i$ form the set $F_i$, also called the {\em local frontier} of $i$. The local frontier of a node evolves dynamically with subproblem branching; its size increases when new subproblems are created and decreases when a subproblem is solved. Since {\em busy nodes} (those with nonempty local frontiers) are able to branch frontier subproblems concurrently, {\em idle nodes} (those with empty local frontiers) request frontier subproblems from another node in order to become busy as fast as possible. This is the basic idea of Algorithm~\ref{alg:karpzhang}. 

For the purpose of handling the partitioning of the subproblems, there are three types of events. In a {\em branching event}, node $i$ performs a branching on a frontier subproblem (lines~\ref{lin:leftmost}--\ref{lin:branching}); in a {\em pairing event}, node $i$ uses a pairing message to request subproblems to some potential donating node (lines~\ref{lin:pairingdest}--\ref{lin:requestingdonation}); and in a {\em donation event}, node $i$ uses a donation message to donate half of its local frontier to a requesting node (lines~\ref{lin:donationset}--\ref{lin:donation}). In every event, node $i$ sends a message to every neighbor in $G$. At least $|N(i)| - 2$ out of these $|N(i)|$ messages are pairing messages without a donation request (line~\ref{lin:notrequestingdonation}). There are two possibilities for the remaining two messages added to $MSG_i$: either both are pairing messages, at most one requesting donation (line~\ref{lin:requestingdonation}) and, consequently, at least one not requesting donation (line~\ref{lin:notrequestingdonation}), or there is a donation message (line~\ref{lin:donation}) and a pairing message (line~\ref{lin:requestingdonation} or~\ref{lin:notrequestingdonation}). For simplicity of presentation, the actions related to termination detection are omitted.

According to line~\ref{lin:leftmost} in Algorithm~\ref{alg:karpzhang}, the frontier subproblem considered in a branching event depends on an ordering of the subproblems from left to right in the search tree. In this ordering, a subproblem $s$ is to the {\em left} of another subproblem $s'$ if $s$ is not in the path from the root to $s'$ and is visited before $s'$ in a depth-first traversal of the search tree. The set of pairs $(i, j)$ such that node $i$ donates to node $j$ is called the {\em pairing set}. For the pair $(i, j)$, the donation event includes, for node $i$, line~\ref{lin:donationset}, where half of the subproblems in $T_i$ are chosen to be transfered to node $j$ ($T_i \subset F_i$ contains the lowest-level subproblems of $F_i$, the {\em level} of a subproblem being the distance from it to the root of the search tree). Initially, $F_i$ gets the initial problem for exactly one node $i$, whereas the others get $\emptyset$. Taking the execution time as the number of branchings in the fully synchronous model~\cite{Barbosa.96}, it is known in the worst case that the Karp-Zhang algorithm is, within constant factors and with high probability, as efficient as any deterministic depth-first algorithm that always chooses the pairing set as large as possible~\cite{Karp.Zhang.93,Ranade.90}.

\begin{algorithm}[htbp]
\caption{Events of an asynchronous version of the Karp-Zhang distributed backtrack search algorithm.}
\label{alg:karpzhang}
\begin{algorithmic}[1]
\Procedure{event$_i$}{$msg_i$, $MSG_i$}
        \State Let $j$ be the origin of $msg_i$
	\State $MSG_i \gets \emptyset$
	\State $N_i \gets N(i)$
	\If{$msg_i$ is a pairing message with a donation request}
		\If{$|F_i| \geq 2$}
			\State Let $D_i \subseteq T_i$ be a set of $\lceil |T_i|/2 \rceil$ subproblems in $T_i$ \label{lin:donationset}
			\State $F_i \gets F_i \setminus D_i$
			\State Add to $MSG_i$ a donation message containing $D_i$ and addressed to $j$ \label{lin:donation}
		\EndIf
		\State $N_i \gets N_i \setminus \{ j \}$
	\ElsIf{$msg_i$ is a donation message}
		\State{Let $D_i$ be the set of subproblems donated by $j$}
		\State $F_i \gets F_i \cup D_i$
	\EndIf
	\If{$F_i \ne \emptyset$}
               	\State Let $s_i$ be the leftmost subproblem in $F_i$ \label{lin:leftmost}
		\State $F_i \gets F_i \setminus \{ s_i \}$
		\State $F_i \gets F_i \cup \{ s \mid s \text{ is a subproblem produced by the branching of } s_i \}$ \label{lin:branching}
	\EndIf
	\If{$F_i = \emptyset$}
		\State $dest_i \gets$ a random member of $N_i$ \label{lin:pairingdest}
		\State $N_i \gets N_i \setminus \{ dest_i \}$
		\State Add to $MSG_i$ a pairing message requesting donation to $dest_i$ \label{lin:requestingdonation}
	\EndIf
	\ForAll{$k \in N_i$}
		\State Add to $MSG_i$ a pairing message not requesting donation to $k$ \label{lin:notrequestingdonation}
	\EndFor
\EndProcedure
\end{algorithmic}
\end{algorithm}

\subsubsection{Partially ordering the messages}

This adaptation of the Karp-Zhang algorithm to our (asynchronous) event-driven model respects the basic conditions for their probabilistic analysis, except for the synchronism and completely connected network assumptions. Relaxing these assumptions in Algorithm~\ref{alg:karpzhang} has two reasons. First, it is more realistic, reducing the communication-related stress of the underlying physical system. Secondly, it avoids most of the synchronization overhead that the synchronous version incurs when the time taken by the branching procedure varies according to the subproblem. The price to pay is, potentially, a less efficient partitioning of the frontier subproblems among the nodes, due to an increase in the probability of unsuccessful donation requests. In order to minimize the effects of this drawback, a dynamic ordering of the messages could be imposed such that donating messages would not be overtaken by sequences of messages of any kind, in particular those ending with a pairing message with a donation request. We give more details in Section~\ref{sec:event}.

\subsection{Iterative methods}
\label{sec:lineareq}

\subsubsection{Systems of linear equations}

Suppose we are given a sparse, invertible $n \times n$ matrix $A$ and a size-$n$ vector $b$ of real numbers. Iterative algorithms that generate a sequence of approximations to the solution vector $x$ of the system of linear equations $Ax = b$ are very efficient methods to solve, in a distributed environment, the systems that arise in a number of engineering and science applications~\cite{Kumar.Grama.Gupta.Karypis.94}. Generally speaking, the sequence starts with an initial guess $x^0$ for the solution vector $x$ and a linear operator is iteratively applied to produce the successive approximations to $x$ until satisfactory convergence is achieved. The linear operator used in each iteration commonly performs a matrix-vector multiplication involving the coefficient matrix $A$ and the approximation vector produced in the previous iteration.

An important issue when distributed implementations of such algorithms are considered is the mutual dependency of the entries of $x$ in the corresponding linear operators, which dramatically affects the convergence speed. Assuming that each node is responsible for computing an entry of the solution vector $x$ and also for the time being that $G$ is completely connected, let $x_i$, for $i \in N$, be the entry of $x$ assigned to node $i$. Since the coefficient matrix $A$ is assumed to be sparse, the matrix-vector multiplication in the linear operator involving $A$ establishes a dependency standard that is related to the nonzero entries of $A$. Let $G_k$ be a subgraph of $G$ defined by the edges that represent the dependency standard for iteration $k$, which means that $ij \in E_k$ if and only if the entry $x_i[k]$ depends on the entry $x_j[k]$ (and conversely); otherwise, either $x_i[k]$ depends on $x_j[k - 1]$ and $x_j[k]$ on $x_i[k - 1]$ or no dependency exists between $x_i$ and $x_j$. The concurrency attainable in an iteration is determined by a {\em coloring of $G_k$}, which turns out to be a mapping $c$ from $N$ to an ordered set of colors such that neighbor nodes get different colors. Note that, for every $i \in N$, the computation carried out in an iteration at every other node getting the same color as $i$ is independent of $x_i$.

Two classical examples of iterative algorithms are the {\em Jacobi} and {\em Gauss-Seidel} methods. For these examples, $G$ is such that its edges represent the nonzero entries of $A$, i.e., $ij \in E$ if and only if the element $a_{ij}$ of $A$ is nonzero. In the Jacobi case, for example, $x_i[k]$ depends on the values produced by the neighbors of $i$ at iteration $k - 1$ in such a way that, for all $k > 0$,
\begin{equation}
x_i[k] = \frac{r_{i, N(i)}[k - 1]}{a_{ii}} + x_i[k - 1],
\label{eq:jacobi}
\end{equation}
where
\begin{equation}
r_{i, N(i)}[k - 1] = b_i - \sum_{j \in N(i)} a_{ij} x_j[k - 1].
\label{eq:residual}
\end{equation}
An execution is said to have converged after iteration $k$ if the magnitude of the {\em residual} $r[k] = b - Ax[k]$ is smaller than a certain tolerance. The quadratic norm is commonly used to compute the magnitude of the residual, and the tolerance is usually a very small number $\epsilon \in (0, 1)$. 

In many situations of practical interest, a coloring can be used to improve the convergence properties of~\eqref{eq:jacobi}, as follows. Let $c$ be a coloring and define $C(i) = N(i) \cap \{ j \mid c(j) < c(i) \}$ (the set of neighbors of $i$ which are assigned colors smaller than $c(i)$) and $\bar C(i) = N(i) \setminus C(i)$. Based on this coloring, the iteration $k$ of a general method becomes
\begin{equation}
x_i[k] = \frac{r_{i, C(i)}[k] + r_{i, \bar C(i)}[k - 1] - b_i}{a_{ii}} + x_i[k - 1],
\label{eq:gauss}
\end{equation}
where $r_{i, C(i)}[k]$ and $r_{i, \bar C(i)}[k - 1]$ are given analogously to~\eqref{eq:residual}. By the definition of $C(i)$, it is clear that the nodes of $G$ operate in the order of their colors during iteration $k$. Notice that~\eqref{eq:jacobi} is obtained from~\eqref{eq:gauss} by simply taking $E_k = \emptyset$ for all $k$ and a coloring that assigns the same color to all nodes (thence $\bar C(i) = N(i)$ for all $i \in N$). On the other hand, the Gauss-Seidel method is obtained by letting $ij \in E_k$ if and only if $a_{ij} \ne 0$.

\subsubsection{Pulse-driven algorithms}

The pulse-driven framework allows the description of the various dependency standards implicit in~\eqref{eq:gauss} solely in terms of the message delivery and pulse generation conditions. By effecting such a separation of computation, communication, and synchronization concerns, one can implement several algorithms with a single implementation of the procedures \Call{event$_i$}{} and \Call{pulse$_i$}{}, provided \Call{getCurrent$_i$}{} and \Call{hasAdvanced$_i$}{} are consistent with~\eqref{eq:gauss} for the coloring at hand. Such flexibility leaves room for several dynamic implementations of~\eqref{eq:gauss}. To be more precise, iteration~\eqref{eq:gauss} can be described by a distributed execution in which the actions performed by the nodes are determined by the procedures in Algorithm~\ref{alg:itlinear}. In this algorithm, node $i$ stores $A_i$ and $b_i$, respectively the $i$th row of $A$ and element of $b$. In addition, it starts out with $k = 0$, $x_i = x_i^0$, $r_i = 0$, and $received_i = |\bar C(i)|$. The number of pulses to compute an iteration is determined by the number of colors used in the coloring, $ncolors$, and the number $nresidual$ of additional pulses required to compute the residual (this latter computation is left unspecified in Algorithm~\ref{alg:itlinear}). The new approximations are computed during the first $ncolors$ pulses, whereas the remaining $nresidual$ pulses are devoted to concluding the computation of the residual.

\begin{algorithm}[htbp]
\caption{Iterative linear operator.}
\label{alg:itlinear}
\begin{algorithmic}[1]
\Procedure{event$_i$}{$msg_i$}
	\State Increment $received_i$ by 1
	\If{$msg_i$ contains $x_j$ for some $j \in N(i)$}
		\State $r_i \gets r_i - a_{ij} x_j$
	\Else
		\State Update residual
	\EndIf
\EndProcedure
\Statex
\Procedure{pulse$_i$}{$\ell_i$, $MSG_i$}
	\If{$received_i = |N(i)| + nresidual$} \label{lin:ifreceived}
		\If{the magnitude of the residual is not in $(0, \epsilon]$}
			\State $x_i \gets r_i/a_{ii} + x_i$
			\State Add to $MSG_i$ a message containing $x_i$ and addressed to each $j \in N(i)$ \label{lin:addxi}
			\State Add to $MSG_i$ the messages related to the computation of the residual \label{lin:startres}
			\State $k \gets k + 1$
			\State $received_i \gets 0$
			\State $r_i \gets b_i - a_{ii} x_i$
		\EndIf
	\Else
		\State Continue updating the residual \label{lin:contres}
	\EndIf
\EndProcedure
\end{algorithmic}
\end{algorithm}

There are two types of message sent by node $i$ in connection with iteration $k$ and, for each of these two types, a maximum delay is established as a function of the coloring $c$. A message of the first type contains an approximation to $x_i$ and is sent to all $j \in N(i)$, as indicated in line~\ref{lin:addxi}. For each $j$, the delay of this message is at most the number of pulses until $j$ needs $x_i[k]$ to update its own approximation, $x_j$. The maximum delay is then either $c(j) - c(i)$, if $i \in C(j)$, or $ncolors + nresidual$, otherwise. A message of the second type is the one used in the computation of the residual in lines~\ref{lin:startres} and~\ref{lin:contres}. The delay of each of these messages should be set in such a way that the comparison in line~\ref{lin:ifreceived} returns $\mathbf{true}$ within at most $ncolors + nresidual$ pulses.

%% file: event.tex
\section{Event-driven computations}
\label{sec:event}

In this section, we introduce a slight variation of the formalism adopted in~\cite{Barbosa.96} as our framework of event-driven computations. Next, we discuss some message delivery conditions, each of them leading to a particular message ordering, that can be used to reduce the intrinsic nondeterminism of such a framework. All references to procedure \Call{event$_i$}{}, $i \in N$, correspond to the procedure described in Section~\ref{sec:intro} for the event-driven category of distributed computations.

\subsection{Model}

A set of functions \Call{event$_i$}{}, for all $i \in N$, defines what we call an {\em event-driven distributed algorithm} $\Gamma$ on $G$. Let $\Xi = \Xi_1 \cup \cdots \cup \Xi_n$ be the set of events of a particular computation of $\Gamma$, $\Xi_i$ being the set of events occurring at $i \in N$ according to Algorithm~\ref{alg:event-driven}. For $\xi_i \in \Xi_i$, let $msg_i(\xi_i)$ be the input message associated with $\xi_i$ and $MSG_i(\xi_i)$ denote the set of messages generated by the occurrence of $\xi_i$. The sequence of local computations that determines the evolution of the local state of each node $i \in N$ is represented by a total order on $\Xi_i$ or, alternatively, by the function $time_i$, defined such that $time_i(\xi_i) = t_i$ if and only if $\xi_i \in \Xi_i$ is the $t_i$th event that occurs at $i$ (we also sometimes say that the time at which $\xi_i$ occurs at $i$ is $t_i$). Messages in $MSG_i(\xi_i)$ trigger other events at a subset of neighbors of $i$. Event $\xi_i$ depends upon $msg_i(\xi_i)$ and the local state resulting from the previous event at the same node. The only special case is that of $time_i(\xi_i) = 1$, in which case $\xi_i$ depends only on the initial local state. 

The causal dependencies of events in $\Xi$ are formally described by means of the usual ``happened before'' partial order $\leadsto$, defined on $\Xi$ as follows~\cite{Lamport.78}. Let $\xi$ be an event. Write $node(\xi)$ for the node at which $\xi$ occurs. We use $\xi \to \xi'$ to denote that either $node(\xi) = node(\xi') = i$, for some $i \in N$, with $time_i(\xi) = time_i(\xi') - 1$, or $node(\xi) \ne node(\xi')$ and $msg(\xi') \in MSG(\xi)$. Given two events $\xi, \xi' \in \Xi$, they satisfy $\xi \leadsto \xi'$ if and only if there exists a sequence $\xi = \xi_1, \ldots, \xi_t = \xi'$ of events in $\Xi$ such that $\xi_s \to \xi_{s + 1}$, for every $s \in \{ 1, \ldots, t - 1 \}$.

\begin{figure}[t]
\centering
\subfigure[{$pred_j(\xi_i) \ne \xi_j$.}]{\input{succnotpred.pstex_t}} 
\subfigure[{$succ_i(\xi_j) \ne \xi_i$.}]{\input{prednotsucc.pstex_t}} 
\caption{$succ_i(\xi_j) = \xi_i \iff pred_j(\xi_i) = \xi_j$ is false.}
\label{fig:predsucc}
\end{figure}
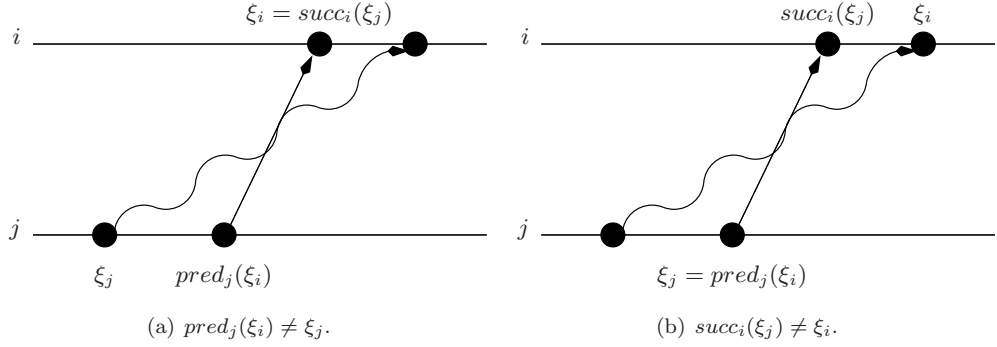

The set $\Xi$ of events also induces the binary relation $pred_j$ (and its analogue $succ_i$) that gives the {\em predecessor} (resp. {\em successor}) at node $j \in N$ (resp. $i \in N$) of an event occurring at node $i \in N$ (resp. $j \in N$)~\cite{Drummond.Barbosa.03}. In more formal terms, given $\xi_i, \xi_j \in \Xi$ such that $node(\xi_j) = j \ne node(\xi_i) = i$, we say that $\xi_j = pred_j(\xi_i)$ if and only if $\xi_j$ is the latest event in $\Xi_j$ such that $\xi_j \leadsto \xi_i$. Note that $pred_j(\xi_i)$ is left undefined if $\xi_j \leadsto \xi_i$ holds for no event $\xi_j \in \Xi_j$. In addition, $time_j(pred_j(\xi_i))$ is assumed to be $0$ in this case. Analogously, $\xi_i = succ_i(\xi_j)$ if and only if $\xi_i$ is the earliest event in $\Xi_i$ such that $\xi_j \leadsto \xi_i$. We leave $succ_i(\xi_j)$ undefined and set $time_i(succ_i(\xi_j)) = 0$ if $\xi_j \leadsto \xi_i$ holds for no event $\xi_i \in \Xi_i$. It should be noted that if $\xi_j \to \xi_i$, then $pred_j(\xi_i) = \xi_j$ and $succ_i(\xi_j) = \xi_i$. However, this may not be the case in the more general situation of $\xi_j \leadsto \xi_i$, as illustrated in Figure~\ref{fig:predsucc}.

Every event $\xi_i \in \Xi_i$, $i \in N$, defines the $n$-dimensional {\em vector clock $E_i(t_i)$}, where $t_i = time_i(\xi_i)$~\cite{Drummond.Barbosa.03,Garg.04}. For $t_i \geq 1$, the $h$th entry of $E_i(t_i)$, for $h \in N$, is given by
\begin{equation}
E_i^h(t_i) = \left\{ \begin{array}{ll}
               t_i,                   & \mbox{if } h = i; \\
	       time_h(pred_h(\xi_i)), & \mbox{otherwise.}
             \end{array} \right.
\label{eq:Vijti}
\end{equation}
The vector clock $E_i(t_i)$ gives, for every node $h \in N$, the time of the latest event $\xi_h$ at $h$ such that $\xi_h \leadsto \xi_i$ (if $h = i$, the current time at $i$), unless no such event exists (and then $E_i^h(t_i) = 0$). For $t_i > 1$, $E_i(t_i)$ is obtained by taking $E_i^h(t_i) = \max \{ E_i^h(t_i - 1), E_j^h(t_j) \}$, where $t_j$ is the time of the event $\xi_j$ at $j \in N(i)$ that originates the message $msg_i(\xi_i)$. A simple way to maintain vector clocks is then to attach $E_j(t_j)$ to every message in $MSG_j(\xi_j)$ upon the occurrence of $\xi_j$. Under the general assumption of an arbitrary message delivery ordering, situations like the ones illustrated in Figure~\ref{fig:predsucc} make such size-$n$ attachments strictly necessary~\cite{Charon-Bost.91}. However, it is possible to use smaller attachments when some specific message delivery ordering is assumed, as we discuss later.

\begin{figure}[t]
\centering
\input{vectorclock.pstex_t}
\caption{The global state induced by the vector clock $E_i(time_i(\xi_i))$.}
\label{fig:vectorclock}
\end{figure}
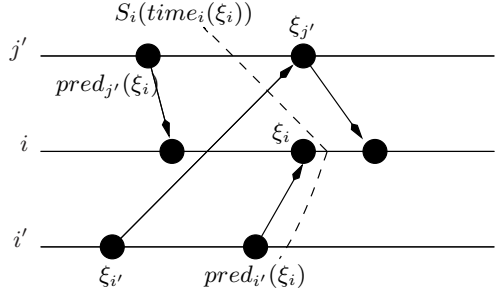

The vector clock $E_i(t_i)$ induces the {\em global state $S_i(t_i)$} with the following characteristics:
\begin{itemize}
\item for each node $j \in N$, $S_i^j(t_i)$ is the local state resulting from the event that occurs at $j$ at time $E_i^j(t_i)$; and

\item for each edge $i'j' \in E$, $S_i^{i' \to j'}(t_i)$ is the set of messages in transit from node $i' \in N$ to node $j' \in N(i')$, i.e., those sent by $i'$ no later than $E_i^{i'}(t_i)$ and received by $j'$ later than $E_i^{j'}(t_i)$; $S_i^{j' \to i'}(t_i)$ is defined similarly.
\end{itemize}
The notions of a vector clock and the global state it induces are depicted in Figure~\ref{fig:vectorclock}. In the global state $S_i(time_i(\xi_i))$, represented in the figure as a dashed line, the events $\xi_i$, $pred_{i'}(\xi_i)$, and $pred_{j'}(\xi_i)$ determine the local states of $i$, $i'$, and $j'$, respectively. The message between $i'$ and $j'$ appears in the state of edge $i'j'$, in the $i' \to j'$ direction.

We henceforth let $M(t_i)$ be the size-$|N(i)|$ vector whose $j$th entry $M_i^j(t_i) = |S_i^{j \to i}(t_i)|$ records the number of messages in transit on edge $ij$ from $j$ to $i$ in global state $S_i(t_i)$.

\subsection{FIFO ordering}

The FIFO message delivery condition is the following: if $\xi_j \to \xi_i$ and $\xi_j' \to \xi_i'$, are two distinct communications between nodes $i \in N$ and $j \in N(i)$, then $\xi_j \leadsto \xi_j' \yields \xi_i \leadsto \xi_i'$ (notice that the two distinct events $\xi_i$ and $\xi_i'$ at the same node $i$ are such that $\xi_i \leadsto \xi_i'$ if and only if $time_i(\xi_i) < time_i(\xi_i')$; the same holds for events $\xi_j$ and $\xi_j'$)~\cite{Garg.04}. This is equivalent to saying that, if $j \in N(i)$ is the origin of the message that triggers the $t_i$th event at node $i$, then no message from $j$ is in transit in $S_i(t_i)$. Formally:
\newcommand{\fdc}{FDC}
\begin{description}
\item[FIFO Delivery Condition (\fdc):] For $i \in N$, $t_i \geq 1$, and $j \in N(i)$ the origin of the message that triggers the $t_i$th event at $i$, $M_i^j(t_i) = 0$.
\end{description}
The implementation of the \fdc\ requires the implementation of the function $M_i^j(t_i)$ for all $t_i \geq 1$. For this purpose, a size-$|N(i)|$ vector $r_i(t_i)$ and a size-$|N(j)|$ vector $s_j(t_j)$ can be used to account for the number of messages exchanged between $i$ and $j$, respectively, and their neighbors. In particular, entry $s_j^i(time_j(\xi_j))$ indicates the number of messages sent by $j$ to $i$ up to, and including, event $\xi_j$. Of these, the number of messages already received by $i$ up to, and including, event $\xi_i$ such that $t_i = time_i(\xi_i)$ is given by $r_i^j(t_i)$~\cite{Garg.04}. Clearly, if $\xi_j$ is the trigger of $\xi_i$, then $s_j^i(time_j(\xi_j)) - r_i^j(t_i)$ messages are in transit from $j$ to $i$ on edge $ij$ in global state $S_i(t_i)$. So $M_i^j(t_i) = s_j^i(time_j(\xi_j)) - r_i^j(t_i)$ and, to allow the checking of whether $M_i^j(t_i) = 0$, node $j$ attaches $s_j^i(time_j(\xi_j))$ to the message sent to $i$ as $\xi_j$ occurs. The use of the improvement for maintaining vector clocks we mentioned earlier, known as the Singhal-Kshemkalyani improvement, is straightforward once the \fdc\ is satisfied~\cite{Singhal.Kshemkalyani.92}.

\subsection{Causal ordering}

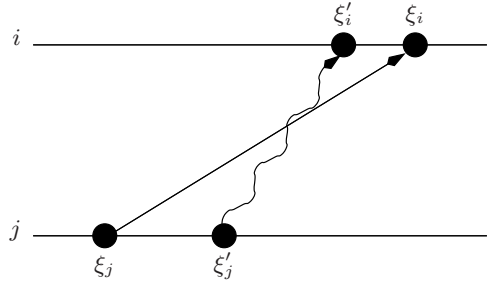
\begin{figure}[t]
\centering
\input{notcausal.pstex_t}
\caption{A single message overtaken by a sequence of messages. The message $\xi_j \to \xi_i$ is in transit in the global state $S_i(time_i(\xi_i'))$.}
\label{fig:notcausal}
\end{figure}

A stronger notion than FIFO ordering is that of {\em causal ordering}, which requires that no single message be overtaken by any sequence of messages~\cite{Birman.Joseph.87,Garg.04,Kshemkalyani.Singhal.98}. Usually, causal ordering is given the following formal statement~\cite{Birman.Joseph.87,Garg.04}. Let $\xi_j \to \xi_i$ be a communication between nodes $j \in N$ and $i \in N(j)$, and $\xi_j' \leadsto \xi_i'$, where $\xi_j' \in \Xi_j$, $\xi_j' \ne \xi_j$, $\xi_i' \in \Xi_j$, and $\xi_i' \ne \xi_i$. Then $\xi_j \leadsto \xi_j' \yields \xi_i \leadsto \xi_i'$. Situations like the one depicted in Figure~\ref{fig:notcausal}, where the last message, say $\xi_k \to \xi_i'$, in the sequence of messages leading from $\xi_j'$ to $\xi_i'$ is such that $\xi_j \leadsto \xi_k$, are not allowed to happen (but notice that situations like those in Figure~\ref{fig:predsucc} may occur even if messages are causally ordered, and therefore $succ_i(\xi_j) = \xi_i \iff pred_j(\xi_i) = \xi_j$ remains false).

The causal ordering is defined in terms of numbers of messages in transit in global states as follows:
\newcommand{\cdc}{CDC}
\begin{description}
\item[Causal Delivery Condition (\cdc):] For $i \in N$, $t_i \geq 1$, and $j \in N(i)$, $M_i^j(t_i) = 0$.
\end{description}
The implementation of the \cdc\ also requires the functions $M_i^j(t_i)$ for all $t_i \geq 1$. But, unlike the case of the \fdc, $M_i^j(t_i)$ is needed whenever $i$ receives a message, not just upon the arrival of a message from $j$. These functions can still be implemented based on $r_i(t_i)$ and $s_j(time_j(pred_j(\xi_i)))$ and the number of messages in transit from $j$ to $i$ on $ij$ in global state $S_i(t_i)$ is still given by $s_j^i(time_j(pred_j(\xi_i))) - r_i^j(t_i)$. But now maintaining $M_i^j(t_i)$ has to be approached similarly to maintaining a vector clock. The only difference in this case is that the attachment received by $i$ along with a message from $k \in N(i)$ represents the view, at $k$, of the number of messages whose sending events by $j$ up to, and including, event $pred_j(\xi_i)$ precede the $t_i$th event at $i$ causally. And since causal ordering implies the \fdc, the simplified implementation of vector clocks mentioned in that case can be easily adapted.

\subsection{Relaxed FIFO ordering and the vector clock algorithm}

In many situations of interest, some degree of nondeterminism is tolerable. In the event-driven framework, a way to introduce some tolerance on message delivery ordering is to allow up to a certain number of messages to be in transit in the global states induced by vector clocks. When applied to \fdc, this relaxation can be formulated as:
\newcommand{\rfdc}{RFDC}
\begin{description}
\item[Relaxed FIFO Delivery Condition (\rfdc):] For $i \in N$, $t_i \geq 1$, and $j \in N(i)$ the origin of the message that triggers the $t_i$th event at $i$, $M_i^j(t_i) \leq \mu_j^i(E_i^j(t_i))$.
\end{description}
In this formulation, $\mu_j^i(E_i^j(t_i))$ is a nonnegative integer function, determined by $j$, that indicates the maximum number of messages which are accepted to be in transit in $S_i(t_i)$ from $j$ to $i$ on $ij$. Clearly, \fdc\ corresponds to \rfdc\ with $\mu_j^i(E_i^j(t_i)) = 0$ for all $t_i \geq 1$. The computation of $M_i^j(t_i)$ can be conducted as before, and \rfdc\ is implemented with the additional attachment of $\mu_j^i(time_j(pred_j(\xi_i)))$ to the message sent to $i$ by $pred_j(\xi_i)$.

One natural question that arises when \rfdc\ is considered is related to its consequences in the computation of vector clocks. According to the main idea in the Singhal-Kshemkalyani vector clock algorithm~\cite{Singhal.Kshemkalyani.92}, and taking into account that the message that triggers $\xi_i$ is allowed to be overtaken by $\mu_j^i(E_i^j(t_i))$ messages sent previously by $j$, node $j$ attaches to that message only the entries of $E_j(time_j(pred_j(\xi_i)))$ that were modified since the earliest of the last $\mu_j^i(E_i^j(t_i))$ transmissions to $i$. For this purpose, node $j$ keeps the additional size-$|N(j)|$ vector $U_j(time_j(\xi_j))$, whose entry $U_j^k(time_j(\xi_j))$ accounts for the number of messages sent to $k \in N(j)$ to which the value of $E_j^k(time_j(\xi_j))$ was attached. Then node $j$ only needs to attach those entries $E_j^k(time_j(\xi_j))$ such that 
\begin{equation}
U_j^k(time_j(\xi_j)) \leq \mu_j^i(E_i^j(t_i)).
\label{eq:update}
\end{equation}

\subsection{Relaxed causal ordering and an algorithm}

The causal ordering defined above may become too strict for a number of communications occurring during an execution, leading to unnecessary loss in concurrency. This fact motivates the definition of a {\em relaxed causal ordering}:
\newcommand{\rcdc}{RCDC}
\begin{description}
\item[Relaxed Causal Delivery Condition (\rcdc):] For $i \in N$, $t_i \geq 1$, and $j \in N(i)$, $M_i^j(t_i) \leq \mu_j^i(E_i^j(t_i))$.
\end{description}
Algorithms~\ref{alg:rcdcsending} and~\ref{alg:rcdcreception} are an implementation of \rcdc\ based on the variables and ideas described earlier. For instance, the vector accounting for the messages sent by each node is maintained by an algorithm similar to that for maintaining the vector clock. The entry of that vector at node $i$ corresponding to nodes $j$ and $k$ is given by $s_j^k(time_j(pred_j(\xi_i)))$ and denoted by $s_i^{j \to k}(t_i)$. The implementation described in Algorithm~\ref{alg:rcdcsending} corresponds to the message sending in lines~\ref{lin:onesend} and~\ref{lin:anothersend} of Algorithm~\ref{alg:event-driven} for node $j$, whereas the reception of a message by node $i$ is implemented in Algorithm~\ref{alg:rcdcreception}. We assume that $t_i$ and $t_j$ are the times of the events that receive and send the messages, respectively. To each message sent by node $j$, it attaches the tolerance $\mu_j^i(t_j)$ and the entries $s_j^{k \to \ell}(t_j)$ chosen so that~\eqref{eq:update} holds, and updates the corresponding $U_j^{k \to \ell}(t_j)$, the entry of the vector $U_j(t_j)$ associated with nodes $k$ and $\ell$. In Algorithms~\ref{alg:rcdcsending} and~\ref{alg:rcdcreception}, \Call{setTolerance$_i$}{} and \Call{getTolerance$_i$}{} are application-specific and are used respectively to attach the tolerance to, or retrieve it from, a message. We give an example next.

\begin{algorithm}[t]
\caption{Implementation of the \rcdc: message sending in lines~\ref{lin:onesend} and~\ref{lin:anothersend} of Algorithm~\ref{alg:event-driven} for node $j$.}
\label{alg:rcdcsending}
\begin{algorithmic}[1]
\Statex {\em Sending, by node $j$, of the messages in $MSG_j(\xi_j)$:}
\ForAll{$i \in N(j)$ for which a message $m$ exists in $MSG_j$}
        \State \Call{setTolerance$_j$}{$m$, $i$}
        \State $\mu_j^i(t_j) \gets$ \Call{getTolerance$_j$}{$m$}
        \State Initialize attachment $a_j^i$ with $\emptyset$
	\ForAll{$k \in N$, $k \ne j$}
		\ForAll{$\ell \in N(k)$}
	                \If {$U_{j, i}^{k \to \ell}(t_j) \leq \mu_j^i(t_j)$}
			        \State Include $s_j^{k \to \ell}(t_j)$ in  $a_j^i$
				\State Increment $U_{j, i}^{k \to \ell}(t_j)$ by 1
	        	\EndIf
		\EndFor
	\EndFor
\EndFor
\ForAll{$m \in MSG_j$}
        \State Let $i$ be the destination of $m$
        \State Increment $s_j^{j \to i}(t_j)$ by 1 and include it in  $a_j^i$
        \State Attach $a_j^i$ to $m$
	\State Send $m$
\EndFor
\end{algorithmic}
\end{algorithm}

\begin{algorithm}[t]
\caption{Implementation of the \rcdc: reception of a message by node $i$.}
\label{alg:rcdcreception}
\begin{algorithmic}[1]
\Statex {\em Upon arrival of message $msg_i$ from node $j$ at node $i$:}
\State Increment $r_i^j(t_i)$ by 1
\State $\mu_j^i \gets$ \Call{getTolerance$_i$}{$msg_i$}
\If{$s_j^{k \to i} - r_i^k(t_i) > \mu_j^i$, for some $k \in N(i)$ such that $s_j^{k \to i}$ is attached to $msg_i$} \label{lin:deliverycond}
        \State Decrement $r_i^j(t_i)$ by 1
        \State Postpone the delivery of $msg_i$
\Else
        \ForAll{$k \in N$, $\ell \in N(k)$ such that $s_j^{k \to \ell}$ is attached to $msg_i$} \label{lin:begindelivery}
                \If {$s_j^{k \to \ell} > s_i^{k \to \ell}(t_i)$}
	                \State $s_i^{k \to \ell}(t_i) \gets s_j^{k \to \ell}$
			\ForAll{$j \in N(i)$}
		                \State $U_{i, j}^{k \to \ell}(t_i) \gets 0$
			\EndFor
       	        \EndIf
        \EndFor
        \State Deliver $msg_i$ \label{lin:enddelivery}
\EndIf
\Statex
\Statex {\em Upon the occurrence of any event at node $i$:}
\While{an undelivered message $msg_i$ exists satisfying the delivery condition of line~\ref{lin:deliverycond}}
       	\State Let $j$ be the origin of $msg_i$
       	\State Increment $r_i^j(t_i)$ by 1
       	\State Execute lines~\ref{lin:begindelivery}--\ref{lin:enddelivery}
\EndWhile
\end{algorithmic}
\end{algorithm}

\subsection{An application: distributed backtrack search}

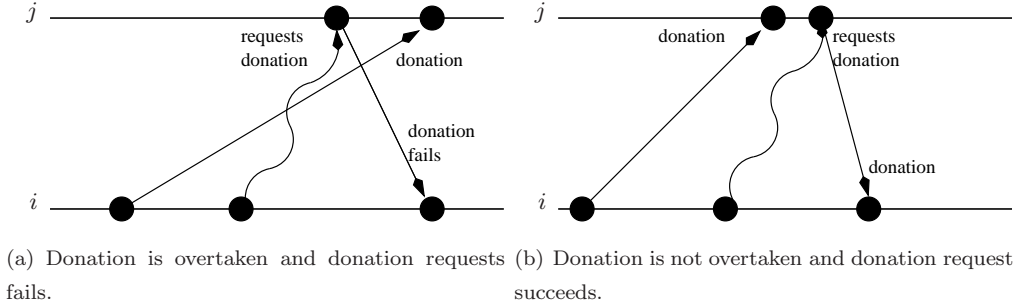
\begin{figure}[t]
\centering
\subfigure[Donation is overtaken and donation requests fails.]{\input{overtaken.pstex_t}} 
\subfigure[Donation is not overtaken and donation request succeeds.]{\input{notovertaken.pstex_t}} 
\caption{Successful response to a donation request may depend on whether it is overtaken by a sequence of messages.}
\label{fig:overtakenornot}
\end{figure}

Let us consider the distributed backtrack search application of Section~\ref{sec:backtrack}. In this application, as indicated before, a dynamic ordering of the messages is useful to avoid the overtaking of donation messages by any message sequences. Recall that there are, besides donation messages, two other types of message in this application, namely pairing messages with and without a donation request. Two situations involving a pairing message with a donation request from $i \in N$ to $j \in N(i)$ and a sequence of messages starting at $i$ and ending by a donation message to $j$ are shown in Figure~\ref{fig:overtakenornot}. In the case in which the donation message is overtaken, the donation request from $i$ fails.

The original idea is then to let the tolerance $\mu_i^j(t_i)$ be determined by node $i$ according to the type of message that is being sent to $j \in N(i)$. If this message is a pairing message with a donation request, then $\mu_i^j(t_i)$ is set to a small value; otherwise, it is set to a very large value. The small value depends on the degree of causality that better suits the current stage of the search. A conservative implementation will choose zero in order to guarantee that donations are never overtaken by donation requests. Algorithm~\ref{alg:tolerance-backtrack} summarizes this.

\begin{algorithm}[t]
\caption{Tolerance setting at node $i$ for the distributed backtrack search.}
\label{alg:tolerance-backtrack}
\begin{algorithmic}[1]
\ForAll{$j \in N(i)$}
        \State $\mu_i^j \gets 0$
\EndFor
\Statex
\Procedure{setTolerance$_i$}{$m$, $j$}
\If{$m$ is a pairing message with a donation request}
        \State $\mu_i^j \gets 0$
\Else
        \State $\mu_i^j \gets \mu_i^j + 1$
\EndIf
\State Attach $\mu_i^j$ to $m$
\EndProcedure
\Statex
\Function{getTolerance$_i$}{$m$}
\State \Return the tolerance attached to $m$
\EndFunction
\end{algorithmic}
\end{algorithm}

%% file: succnotpred.pstex_t
\begin{picture}(0,0)%
\includegraphics{succnotpred.pstex}%
\end{picture}%
\setlength{\unitlength}{3947sp}%
\begingroup\makeatletter\ifx\SetFigFont\undefined%
\gdef\SetFigFont#1#2#3#4#5{%
  \reset@font\fontsize{#1}{#2pt}%
  \fontfamily{#3}\fontseries{#4}\fontshape{#5}%
  \selectfont}%
\fi\endgroup%
\begin{picture}(3137,1831)(1826,-3410)
\put(2026,-1861){\makebox(0,0)[rb]{\smash{{\SetFigFont{9}{10.8}{\rmdefault}{\mddefault}{\updefault}$i$}}}}
\put(2026,-3061){\makebox(0,0)[rb]{\smash{{\SetFigFont{9}{10.8}{\rmdefault}{\mddefault}{\updefault}$j$}}}}
\put(2551,-3361){\makebox(0,0)[b]{\smash{{\SetFigFont{9}{10.8}{\rmdefault}{\mddefault}{\updefault}$\xi_j$}}}}
\put(3301,-3361){\makebox(0,0)[b]{\smash{{\SetFigFont{9}{10.8}{\rmdefault}{\mddefault}{\updefault}$pred_j(\xi_i)$}}}}
\put(3901,-1711){\makebox(0,0)[b]{\smash{{\SetFigFont{9}{10.8}{\rmdefault}{\mddefault}{\updefault}$\xi_i = succ_i(\xi_j)$}}}}
\end{picture}%

%% file: prednotsucc.pstex_t
\begin{picture}(0,0)%
\includegraphics{prednotsucc.pstex}%
\end{picture}%
\setlength{\unitlength}{3947sp}%
\begingroup\makeatletter\ifx\SetFigFont\undefined%
\gdef\SetFigFont#1#2#3#4#5{%
  \reset@font\fontsize{#1}{#2pt}%
  \fontfamily{#3}\fontseries{#4}\fontshape{#5}%
  \selectfont}%
\fi\endgroup%
\begin{picture}(3137,1831)(6026,-3410)
\put(6226,-1861){\makebox(0,0)[rb]{\smash{{\SetFigFont{9}{10.8}{\rmdefault}{\mddefault}{\updefault}$i$}}}}
\put(6226,-3061){\makebox(0,0)[rb]{\smash{{\SetFigFont{9}{10.8}{\rmdefault}{\mddefault}{\updefault}$j$}}}}
\put(8701,-1711){\makebox(0,0)[b]{\smash{{\SetFigFont{9}{10.8}{\rmdefault}{\mddefault}{\updefault}$\xi_i$}}}}
\put(7501,-3361){\makebox(0,0)[b]{\smash{{\SetFigFont{9}{10.8}{\rmdefault}{\mddefault}{\updefault}$\xi_j = pred_j(\xi_i)$}}}}
\put(8101,-1711){\makebox(0,0)[b]{\smash{{\SetFigFont{9}{10.8}{\rmdefault}{\mddefault}{\updefault}$succ_i(\xi_j)$}}}}
\end{picture}%

%% file: vectorclock.pstex_t
\begin{picture}(0,0)%
\includegraphics{vectorclock.pstex}%
\end{picture}%
\setlength{\unitlength}{3947sp}%
\begingroup\makeatletter\ifx\SetFigFont\undefined%
\gdef\SetFigFont#1#2#3#4#5{%
  \reset@font\fontsize{#1}{#2pt}%
  \fontfamily{#3}\fontseries{#4}\fontshape{#5}%
  \selectfont}%
\fi\endgroup%
\begin{picture}(3168,1831)(1645,-3935)
\put(1876,-2461){\makebox(0,0)[rb]{\smash{{\SetFigFont{9}{10.8}{\rmdefault}{\mddefault}{\updefault}$j'$}}}}
\put(1876,-3661){\makebox(0,0)[rb]{\smash{{\SetFigFont{9}{10.8}{\rmdefault}{\mddefault}{\updefault}$i'$}}}}
\put(1876,-3061){\makebox(0,0)[rb]{\smash{{\SetFigFont{9}{10.8}{\rmdefault}{\mddefault}{\updefault}$i$}}}}
\put(3526,-2986){\makebox(0,0)[rb]{\smash{{\SetFigFont{9}{10.8}{\rmdefault}{\mddefault}{\updefault}{\color[rgb]{0,0,0}$\xi_i$}%
}}}}
\put(3601,-2311){\makebox(0,0)[b]{\smash{{\SetFigFont{9}{10.8}{\rmdefault}{\mddefault}{\updefault}{\color[rgb]{0,0,0}$\xi_{j'}$}%
}}}}
\put(2701,-2686){\makebox(0,0)[rb]{\smash{{\SetFigFont{9}{10.8}{\rmdefault}{\mddefault}{\updefault}{\color[rgb]{0,0,0}$pred_{j'}(\xi_i)$}%
}}}}
\put(2401,-3886){\makebox(0,0)[b]{\smash{{\SetFigFont{9}{10.8}{\rmdefault}{\mddefault}{\updefault}{\color[rgb]{0,0,0}$\xi_{i'}$}%
}}}}
\put(3301,-3886){\makebox(0,0)[b]{\smash{{\SetFigFont{9}{10.8}{\rmdefault}{\mddefault}{\updefault}{\color[rgb]{0,0,0}$pred_{i'}(\xi_i)$}%
}}}}
\put(2851,-2236){\makebox(0,0)[b]{\smash{{\SetFigFont{9}{10.8}{\rmdefault}{\mddefault}{\updefault}{\color[rgb]{0,0,0}$S_i(time_i(\xi_i))$}%
}}}}
\end{picture}%

%% file: notcausal.pstex_t
\begin{picture}(0,0)%
\includegraphics{notcausal.pstex}%
\end{picture}%
\setlength{\unitlength}{3947sp}%
\begingroup\makeatletter\ifx\SetFigFont\undefined%
\gdef\SetFigFont#1#2#3#4#5{%
  \reset@font\fontsize{#1}{#2pt}%
  \fontfamily{#3}\fontseries{#4}\fontshape{#5}%
  \selectfont}%
\fi\endgroup%
\begin{picture}(3137,1756)(1826,-3335)
\put(2026,-1861){\makebox(0,0)[rb]{\smash{{\SetFigFont{9}{10.8}{\rmdefault}{\mddefault}{\updefault}$i$}}}}
\put(2026,-3061){\makebox(0,0)[rb]{\smash{{\SetFigFont{9}{10.8}{\rmdefault}{\mddefault}{\updefault}$j$}}}}
\put(4501,-1711){\makebox(0,0)[b]{\smash{{\SetFigFont{9}{10.8}{\rmdefault}{\mddefault}{\updefault}$\xi_i$}}}}
\put(4051,-1711){\makebox(0,0)[b]{\smash{{\SetFigFont{9}{10.8}{\rmdefault}{\mddefault}{\updefault}$\xi_i'$}}}}
\put(2551,-3286){\makebox(0,0)[b]{\smash{{\SetFigFont{9}{10.8}{\rmdefault}{\mddefault}{\updefault}$\xi_j$}}}}
\put(3301,-3286){\makebox(0,0)[b]{\smash{{\SetFigFont{9}{10.8}{\rmdefault}{\mddefault}{\updefault}$\xi'_j$}}}}
\end{picture}%

%% file: overtaken.pstex_t
\begin{picture}(0,0)%
\includegraphics{overtaken.pstex}%
\end{picture}%
\setlength{\unitlength}{3947sp}%
\begingroup\makeatletter\ifx\SetFigFont\undefined%
\gdef\SetFigFont#1#2#3#4#5{%
  \reset@font\fontsize{#1}{#2pt}%
  \fontfamily{#3}\fontseries{#4}\fontshape{#5}%
  \selectfont}%
\fi\endgroup%
\begin{picture}(3137,1403)(1826,-3144)
\put(2026,-1861){\makebox(0,0)[rb]{\smash{{\SetFigFont{9}{10.8}{\rmdefault}{\mddefault}{\updefault}$j$}}}}
\put(2026,-3061){\makebox(0,0)[rb]{\smash{{\SetFigFont{9}{10.8}{\rmdefault}{\mddefault}{\updefault}$i$}}}}
\end{picture}%

%% file: notovertaken.pstex_t
\begin{picture}(0,0)%
\includegraphics{notovertaken.pstex}%
\end{picture}%
\setlength{\unitlength}{3947sp}%
\begingroup\makeatletter\ifx\SetFigFont\undefined%
\gdef\SetFigFont#1#2#3#4#5{%
  \reset@font\fontsize{#1}{#2pt}%
  \fontfamily{#3}\fontseries{#4}\fontshape{#5}%
  \selectfont}%
\fi\endgroup%
\begin{picture}(3137,1403)(1826,-3144)
\put(2026,-1861){\makebox(0,0)[rb]{\smash{{\SetFigFont{9}{10.8}{\rmdefault}{\mddefault}{\updefault}$j$}}}}
\put(2026,-3061){\makebox(0,0)[rb]{\smash{{\SetFigFont{9}{10.8}{\rmdefault}{\mddefault}{\updefault}$i$}}}}
\end{picture}%

%% file: pulse.tex
\section{Pulse-driven computations}
\label{sec:pulse}

A {\em pulse-driven distributed algorithm} is defined by a set of functions \Call{pulse$_i$}{} and \Call{event$_i$}{}, for all $i \in N$. Once again, for given initial local states, there are multiple possible executions of a pulse-driven distributed algorithm. There are two sources for this nondeterminism: the delays between consecutive pulses at a single node and, as in the event-driven framework, those the messages undergo to be delivered. In this context, an appropriate ordering of pulses and messages reduces the nondeterminism of a pulse-driven algorithm.

\subsection{Model}

In addition to the set $\Xi$ of events, what characterizes a pulse-driven computation on $G$ is a set $\Lambda = \Lambda_1 \cup \cdots \cup \Lambda_n$ of {\em pulses}. For each $i \in N$, $\Lambda_i$ stands for the pulses that occur at node $i$. For $\lambda_i \in \Lambda_i$, we let $MSG_i(\lambda_i)$ denote the set of messages generated by the occurrence of $\lambda_i$, represented in Algorithm~\ref{alg:pulse-driven} by the execution of \Call{pulse$_i$}{}. The function $rank_i$ is used to order the pulses at node $i \in N$ and returns the rank of each pulse at $i$. The procedure \Call{pulse$_i$}{} has $rank_i(\lambda_i)$ as input and each of the resulting messages (those in $MSG_i(\lambda_i)$) triggers an event associated with some pulse at a neighbor of $i$.

In this framework for describing and analyzing pulse-driven computations, the local state updates of any node are driven by a {\em local clock mechanism}, which determines the execution of the pulses as described in Algorithm~\ref{alg:pulse-driven}. Particularly important is how the message-triggered events in $\Xi_i$ are related to the pulses in $\Lambda_i$. We assume that every event $\xi_i \in \Xi_i$ is associated with a pulse $\lambda_i \in \Lambda_i$ via a function $pulse_i$ such that $\lambda_i = pulse_i(\xi_i)$. This means that $\xi_i$ occurs at $i$ after $i$'s local clock generates the $(\ell_i - 1)$th pulse, where $\ell_i = rank_i(\lambda_i)$, and before pulse $\lambda_i$ is generated. Conversely, with every pulse $\lambda_i \in \Lambda_i$ is associated a set of events, possibly empty, consisting of the intermediate computations occurring at $i$ between the generation of the $(\ell_i - 1)$th pulse and the $\ell_i$th, where $\ell_i = rank_i(\lambda_i)$.

\begin{figure}[t]
\centering
\input{asexecution.pstex_t}
\caption{Pulse-driven execution on a distributed system.}
\label{fig:asexecution}
\end{figure}
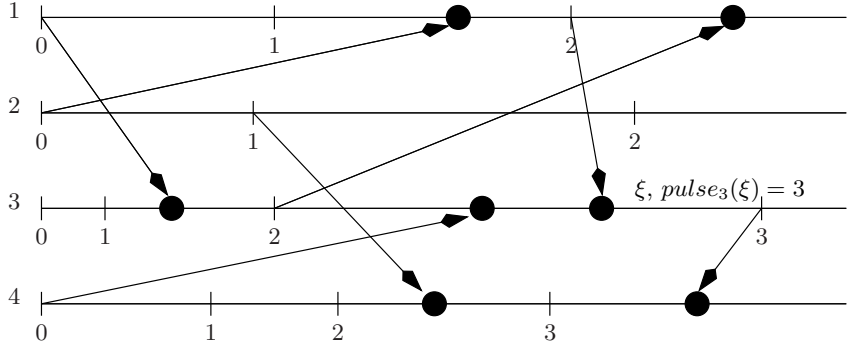

The relation between events at distinct nodes, illustrated in Figure~\ref{fig:asexecution}, is formalized by a re-definition of the notation established for the precedence between events. For two distinct events $\xi_i$ and $\xi_i'$ at a single node $i \in N$, $\xi_i \to \xi_i'$ indicates that $\xi_i$ is the latest event occurring at $i$ such that $rank_i(pulse_i(\xi_i)) < rank_i(pulse(\xi_i'))$ (note that events at a same pulse are unrelated). Now let $\xi_i$ and $\xi_j$ be two events at nodes $i$ and $j \in N(i)$, respectively. The notation $\xi_i \to \xi_j$ is re-defined in the pulse-driven context to indicate that $msg(\xi_j) \in MSG(\lambda_i)$, where $\lambda_i = pulse_i(\xi_i)$. This condition says that $i$'s local state due to the occurrence of $\lambda_i$ influences the local state of $j$ due to pulse $\lambda_j = pulse_j(\xi_j)$.

The precedence relation between the pulses of a pulse-driven computation is not determined by event-triggering messages but is, instead, defined on the number of pulses generated by each local clock mechanism. Naturally, if $\lambda_i$ and $\lambda_i'$ are two pulses at $i$, then we write $\lambda_i \to \lambda_i'$ to denote $rank_i(\lambda_i) = rank_i(\lambda_i') - 1$. We extend this notion almost directly to the case of pulses at different nodes: given $\lambda_i \in \Lambda_i$ and $\lambda_j \in \Lambda_j$ such that $rank_j(\lambda_j) > 1$, $j \in N(i)$, we write $\lambda_i \to \lambda_j$ to mean that $rank_i(\lambda_i) = rank_j(\lambda_j) - 1$.

The relation $\leadsto$ is defined between events or between pulses based on the appropriate $\to$ like in the event-driven framework. We say that $\lambda_i = pred_i(\lambda_j)$, for $i, j \in N$, if $\lambda_i$ is the latest pulse at $i$ such that $\lambda_i \leadsto \lambda_j$. Note that $rank_i(pred_i(\lambda_j)) = rank_j(\lambda_j) - 1$ for $j \in N(i)$, provided $rank_j(\lambda_j) > 1$ (otherwise, $pred_j(\lambda_j)$ does not exist and we assume $rank_i(pred_i(\lambda_j)) = 0$). More generally, $rank_j(\lambda_j) - rank_i(pred_i(\lambda_j))$ is the distance between $i$ and $j$ in $G$ so long as $pred_i(\lambda_j)$ exists (otherwise, $rank_i(pred_i(\lambda_j)) = 0$). The {\em vector clock $P_i(\ell_i)$} of node $i$ at pulse $\lambda_i$ is defined as the vector whose $j$th entry, for $j \in N$, is given by
\begin{equation}
P_i^j(\ell_i) = \left\{ \begin{array}{ll}
                 \ell_i,                    & \mbox{if } j = i; \\
	         rank_j(pred_j(\lambda_i)), & \mbox{otherwise,}
                \end{array} \right.
\label{eq:Vijli}
\end{equation}
where $\ell_i = rank_i(\lambda_i)$. The definition of the {\em global state $S_i(\ell_i)$} associated with vector clock $P_i(\ell_i)$ is straightforward, as follows:
\begin{itemize}
\item for each node $j \in N$, $S_i^j(\ell_i)$ is the local state resulting from the pulse at $i$ whose rank is $P_i^j(\ell_i)$; and

\item for each edge $i'j' \in E$, $S_i^{i' \to j'}(\ell_i)$ is the set of messages in transit from node $i' \in N$ to node $j' \in N(i')$, i.e., those sent by $i$ at a pulse ranking no more than $P_i^{i'}(\ell_i)$ and received by $j'$ through event $\xi_j$ such that $pulse_j(\xi_j)$ ranks more than $P_i^{j'}(\ell_i)$; $S_i^{j' \to i'}(\ell_i)$ is defined similarly.
\end{itemize}

\subsection{Synchronous ordering}

The strongest pulse and message ordering one may define is the {\em synchronous ordering}. This ordering requires that the delay a message from pulse $\lambda_i$ at node $i \in N$ undergoes to be delivered to $i$'s neighbor $j$ be bounded by the local times at $j$ at which the $(\ell_i = rank_i(\lambda_i))$th and $(\ell_i + 1)$th pulses occur~\cite{Barbosa.96}. Put differently, the synchronous ordering imposes an order on the events with respect to pulses. Let $\lambda_j$ be a pulse at $j$ and let $\ell_j = rank_j(\lambda_j)$. Then messages are ordered such that $\xi_i \to \xi_j$ only if $\lambda_i \to \lambda_j$ (i.e., $\ell_j - \ell_i = 1$), where $\xi_i$ and $\xi_j$ are events at $i$ and $j$, respectively, with $pulse_i(\xi_i) = \lambda_i$ and $pulse_j(\xi_j) = \lambda_j$. What is meant here is that, by hypothesis, as many pulses as $\ell_i$ must have been performed so far at $j$ when a message in $MSG_i(\lambda_i)$ triggers an event at $j$. Under these conditions, the absence of messages between pulses $\lambda_i$ and $\lambda_j$ when $\lambda_i \to \lambda_j$ provides information to node $j$: the local state of $i$ resulting from the occurrence of $\lambda_i$ is irrelevant to the occurrence of $\lambda_j$.

A necessary condition for the synchronous ordering establishes the number of pulses that are required to occur before the reception of each message, as follows:
\newcommand{\sdc}{SDC}
\begin{description}
\item[Synchronous Delivery Condition (\sdc):] For $i \in N$, $j \in N(i)$, and $\xi_j$ the reception at $j$ of a message sent by pulse $\lambda_i$,  $rank_j(pulse_j(\xi_j)) \geq rank_i(\lambda_i) + 1$.
\end{description}
In addition to the postponing of message deliveries the \sdc\ may cause, the synchronous ordering also requires that some conditions be satisfied for pulse $\lambda_i$ to take place at $i \in N$. This stems from the fact that the occurrence of $\lambda_i$ depends on the messages sent to $i$ from the pulses at $i$'s neighbors that rank less than $\lambda_i$. Pulse $\lambda_i$ may only occur at $i$ when the following condition holds:
\newcommand{\spgc}{SPGC}
\begin{description}
\item[Synchronous Pulse Generation Condition (\spgc):] For $i \in N$ and $j \in N(i)$, $S_i^{j \to i}(rank_i(\lambda_i)) = \emptyset$.
\end{description}
A natural question to ask is whether the \sdc\ and the \spgc\ induce the synchronous ordering as claimed. To see that this is indeed the case, consider pulse $\lambda_i$ at node $i \in N$ and let $msg_j \in MSG_i(\lambda_i)$ trigger event $\xi_j$ at $j \in N(i)$. By the \sdc, $rank_j(\lambda_j) \geq rank_i(\lambda_i) + 1$, with $\lambda_j = pulse_j(\xi_j)$. In addition, by the \spgc, $msg_j$ must not be in transit in $S_j(rank_i(\lambda_i) + 1)$, which leads to $rank_j(\lambda_j) \leq rank_i(\lambda_i) + 1$. Figure~\ref{fig:violatedsdcspgc} is an illustration of the situations that are forbidden by the \sdc\ and the \spgc, with $\ell_i = rank_i(\lambda_i)$ and $\ell_j = rank_j(\lambda_j)$. Implementations of the synchronous ordering are the celebrated $\alpha$, $\beta$, and $\gamma$ synchronizers~\cite{Barbosa.96,Garg.04}.

\begin{figure}[t]
\centering
\subfigure[{The \sdc\ is violated.}]{\input{violatedsdc.pstex_t}} 
\subfigure[{The \spgc\ is violated.}]{\input{violatedspgc.pstex_t}} 
\caption{Two scenarios forbidden by the synchronous ordering.}
\label{fig:violatedsdcspgc}
\end{figure}
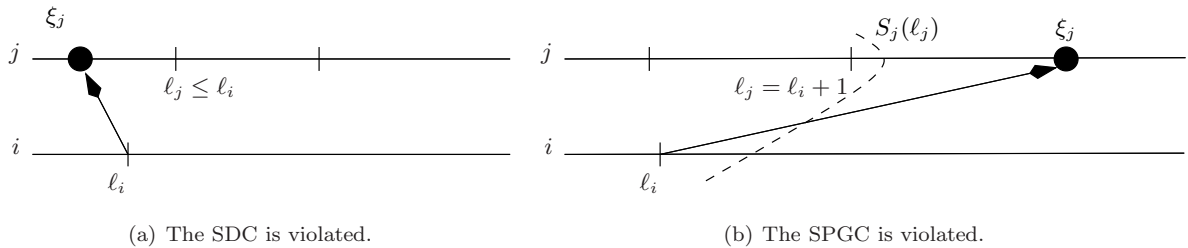

An additional observation in connection with the synchronous ordering is that it also leads to a causal ordering of the messages in the sense of the re-definition of the precedence relation between events. It is a simple matter to check that, if a sequence of messages overtakes a single message, then either the \sdc\ or the \spgc\ is violated. If the \sdc\ is not violated, then the last message of the sequence is received at least two pulses after the pulse that originates the first message of the sequence (since the sequence has length at least 2). This means that the single message is received at least three pulses later, which violates the \spgc.

\subsection{Partially synchronized ordering and an algorithm}

The relaxation of the synchronous ordering is called the {\em partially synchronous ordering} and is defined through relaxations of the \sdc\ and the \spgc. The delay of a message is bounded from below and above by integer functions, which affects the reception of messages and the generation of pulses as follows:
\newcommand{\psdc}{PSDC}
\begin{description}
\item[Partially Synchronous Delivery Condition (\psdc):] For $i \in N$, $j \in N(i)$, $\xi_j$ the reception at $j$ of a message sent by pulse $\lambda_i$, and $\rho_i^j(rank_i(\lambda_i)) \geq 1$, $rank_j(pulse_j(\xi_j)) \geq rank_i(\lambda_i) + \rho_i^j(rank_i(\lambda_i))$.
\end{description}
\newcommand{\pspgc}{PSPGC}
\begin{description}
\item[Partially Synchronous Pulse Generation Condition (\pspgc):] For $i \in N$, $j \in N(i)$, and $0 \leq \delta_i^j(rank_i(\lambda_i)) \leq rank_i(\lambda_i)$, $S_i^{j \to i}(rank_i(\lambda_i) - \delta_i^j(rank_i(\lambda_i))) \cap S_i^{j \to i}(rank_i(\lambda_i)) = \emptyset$.
\end{description}
The lower and upper bounds for the delay of a message are given in the \psdc\ and the \pspgc, respectively, by $\rho_i^j(rank_i(\lambda_i))$ and $\delta_i^j(rank_i(\lambda_i))$. In the \psdc, a message from $\lambda_i$ is to be received by $j$ only after the occurrence of a number of pulses at $j$ greater than $rank_i(\lambda_i)$ by a value determined by $i$ at pulse $\lambda_i$. In the \pspgc, node $i$ determines, for $\lambda_i$, a rank at $j$ such that all messages sent by $j$ to $i$ before, and including, the pulse of this rank must have been received before the occurrence of $\lambda_i$.

Some preliminary observations with respect to the combined implementation of the \psdc\ and the \pspgc\ are as follows. Consider pulse $\lambda_i$ and let $\ell_i = rank_i(\lambda_i)$ at node $i \in N$. Every message sent by $\lambda_i$ to some $j \in N(i)$ carries $\ell_i$ and $\rho_i^j(\ell_i)$ attached to it. A message $msg_i$ from node $j \in N(i)$ is accepted only when the pulse rank $\ell_j$ attached to $msg_i$ is at most $\ell_i - \rho_j^i(\ell_j)$. Consequently, if $msg_i$ arrives at $i$ when $\ell_i < \ell_j + \rho_j^i(\ell_j)$, then $i$ must wait until pulse $\ell_j + \rho_j^i(\ell_j)$ occurs before accepting $msg_i$. For this reason, $msg_i$ can only be accepted by $i$ if a pulse ranking $\ell_i$ exists such that both $\ell_i - \delta_i^j(\ell_i) \leq \ell_j$ and $\ell_i \geq \ell_j + \rho_j^i(\ell_j)$ or, equivalently, such that $\rho_j^i(\ell_j) \leq \ell_i - \ell_j \leq \delta_i^j(\ell_i)$. Otherwise, there is no pulse at $i$ at which the reception of $msg_i$ satisfies both the \psdc\ and the \pspgc. Therefore, one difficulty of implementing a partially synchronous ordering is that the functions $\rho_j^i$, determined by $j$, and $\delta_i^j$, determined by $i$, must be compatible in the sense that there is a pulse at $i$, for every message sent by $j$, at which it can be accepted. We henceforth assume that this is the case.

In addition to message postponing, the implementation of the partially synchronous ordering involves the control of pulse occurrences. In this context, one additional difficulty related to the combined implementation of the \psdc\ and the \pspgc\ is that of handling the ``absence of messages'' between two pulses. The reason for this is that the occurrence of $\lambda_i$ depends on the number of messages sent to $i$ from certain specific pulses at $i$'s neighbors. Pulse $\lambda_i$ only occurs after every message sent by $j \in N(i)$ to $i$ in connection with the pulse of rank $\ell_i - \delta_i^j(\ell_i)$ triggers an event at $i$.

\begin{algorithm}[tp]
\caption{Implementation of the \psdc\ and the \pspgc: message sending in lines~\ref{lin:onesend2} and~\ref{lin:anothersend2} of Algorithm~\ref{alg:pulse-driven} for node $j$, reception of a message by node $i$, and the local clock mechanism of node $i$.}
\label{alg:pspgc}
\begin{algorithmic}[1]
\Statex {\em Sending, by node $j$, of the messages in $MSG_j(\lambda_j)$:}
\ForAll{$i \in N(j)$}
        \State Initialize control message $m$ with $\ell_j$ and $|MSG_j^i|$
	\State Send $m$
\EndFor
\ForAll{message $msg_j \in MSG_j^i$}
        \State Attach $\ell_j$ to $msg_j$
	\State \Call{setMinimumDelay$_j$}{$msg_j$, $\ell_j$, $i$}
	\State Send $msg_j$
\EndFor
\Statex
\Function{hasAdvanced$_i$}{}
\While{a control message $m$ from $j \in N(i)$ exists with attachments $\ell_j$ and $|MSG_j^i|$}
        \State $safe_i^j(\ell_j) \gets \mathbf{true}$
        \State Increment $pending_i^j(\ell_j)$ by $|MSG_j^i|$
\EndWhile
\If{$safe_i^j(\ell_i - \delta_i^j(\ell_i)) \mathbf{\ and\ } pending_i^j(\ell_i - \delta_i^j(\ell_i)) = 0$ for all $j \in N(i)$}
	\State Increment $\ell_i$ by 1
	\State \Return $\mathbf{true}$
\EndIf
\State \Return $\mathbf{false}$
\EndFunction
\Statex
\Statex {\em Upon arrival of message $msg_i$ from node $j$ at node $i$ with attachment $\ell_j$:}
\State $\rho_j^i \gets$ \Call{getMinimumDelay$_i$}{$msg_i$} \label{lin:begindelivery2}
\If{$\ell_i \geq \ell_j + \rho_j^i$} \label{lin:deliverycond2}
	\State Decrement $pending_i^j(\ell_j)$ by 1
	\State Deliver $msg_i$ \label{lin:enddelivery2}
\Else
	\State Postpone the delivery of $msg_i$
\EndIf
\Statex
\Statex {\em Upon the occurrence of any pulse at node $i$:}
\While{an undelivered message $msg_i$ exists satisfying the delivery condition of line~\ref{lin:deliverycond2}}
       	\State Execute lines~\ref{lin:begindelivery2}--\ref{lin:enddelivery2}
\EndWhile
\end{algorithmic}
\end{algorithm}

Some of the variables, messages, and computation associated with Algorithm~\ref{alg:pspgc} are related to the control of the execution. A variable $\ell_i$ is used to implement $i$'s local clock mechanism. Its initial value is 0 to indicate that the first pulse at node $i$ has not yet occurred. The subset of $MSG_i(\lambda_i)$, where $\lambda_i \in \Lambda_i$, constituted by the messages addressed to $j \in N(i)$ is represented simply by $MSG_i^j$. Two additional control messages are used: $safe_i^j(\ell_i)$ stores a Boolean value indicating whether the control message associated with pulse $\lambda_j$ at $j \in N(i)$ such that $\ell_j = rank_j(\lambda_j)$ has been received by $i$; $pending_i^j(\ell_j)$, whose initial value is 0, is the number of messages sent by $\lambda_j$ and not yet received by $i$. The control messages affect the ordering of the computation, which in turn may change the state of the control variables. This depends upon the minimum delay $\rho_j^i(rank_j(\lambda_j))$ attached to every message sent by $\lambda_j$ by the application-specific procedure \Call{setMinimumDelay$_j$}{} and retrieved by $i$ via function \Call{getMinimumDelay$_i$}{}. The implementation of function \Call{getCurrent$_i$}{} of Section~\ref{sec:intro} is not shown in Algorithm~\ref{alg:pspgc}, since it simply returns $\ell_i$.

\subsection{An application: systems of linear equations}

Let us return to the iterative algorithm for systems of linear equations described in Section~\ref{sec:lineareq}. Recall that $ncolors + nresidual$ pulses occur at $i \in N$ at each iteration $k$ of this algorithm, and that the computation of $x_i[k]$ is accomplished as soon as all the information it needs from its neighbors is available. The minimum and maximum delays to be used with the partially synchronized ordering of Algorithm~\ref{alg:pspgc} are set as shown in Algorithm~\ref{alg:itlineardelay} in order to ensure that $x_i[k]$ is computed in one of the first $ncolors$ pulses of iteration $k$ without interfering with the computation of the residual. These delays are determined based on a coloring in which the colors are numbered from the set $\{ 0, 1, \ldots, ncolors - 1 \}$ and are such that the computations at an iteration are performed in increasing order of the colors in this set. Let $dist(a, b)$ be the ``circular distance'' between two colors $a$ and $b$, given by $b - a$, if $b \geq a$, or $b + ncolors - a$, otherwise. At iteration $k$, the maximum delay $\delta_i^j(k(ncolors + nresidual) + c(i))$, for $j \in N(i)$, is determined so that the computation of $x_i[k]$ is accomplished in a pulse no later than the one ranked $k(ncolors + nresidual) + c(i)$. The maximum delay is then given by $dist(c(j), c(i))$ for this pulse, which ensures that $x_j[k]$, if $j \in C(i)$, and $x_j[k - 1]$, if $j \in \bar C(i)$, are available to $i$ when needed at iteration $k$. The minimum delay is set in procedure \Call{setMinimumDelay$_i$}{} of Algorithm~\ref{alg:itlineardelay} in such a way that $x_i[k]$ is received by $j \in N(i)$ such that $c(j) < c(i)$ in a pulse no earlier than the first pulse of iteration $k + 1$, which only occurs after the pulses devoted to the computation of the residual of iteration $k$.

\begin{algorithm}[t]
\caption{Delays at node $i$, for $j \in N(i)$ and the colors $c(i)$ and $c(j)$.}
\label{alg:itlineardelay}
\begin{algorithmic}[1]
\If{$\ell_i = k(ncolors + nresidual) + c(i)$}
	\If{$c(j) < c(i)$}
		\State $\delta_i^j(\ell_i) \gets dist(c(j), c(i))$
	\Else
		\State $\delta_i^j(\ell_i) \gets dist(c(j), c(i)) + nresidual$
	\EndIf
\Else
	\State Increment $\delta_i^j(\ell_i)$ by 1
\EndIf
\Statex
\Procedure{setMinimumDelay$_i$}{$m$, $\ell_i$, $j$}
\If{$c(j) > c(i)$}
        \State Attach 1 to $m$
\Else
        \State Attach $ncolors + nresidual - \ell_i \mod (ncolors + nresidual)$ to $m$
\EndIf
\EndProcedure
\Statex
\Function{getMinimumDelay$_i$}{$m$}
\State \Return the minimum delay attached to $m$
\EndFunction
\end{algorithmic}
\end{algorithm}

%% file: asexecution.pstex_t
\begin{picture}(0,0)%
\includegraphics{asexecution.pstex}%
\end{picture}%
\setlength{\unitlength}{3947sp}%
\begingroup\makeatletter\ifx\SetFigFont\undefined%
\gdef\SetFigFont#1#2#3#4#5{%
  \reset@font\fontsize{#1}{#2pt}%
  \fontfamily{#3}\fontseries{#4}\fontshape{#5}%
  \selectfont}%
\fi\endgroup%
\begin{picture}(5431,2173)(1969,-3926)
\put(3796,-3286){\makebox(0,0)[b]{\smash{{\SetFigFont{9}{10.8}{\rmdefault}{\mddefault}{\updefault}$2$}}}}
\put(5526,-3886){\makebox(0,0)[b]{\smash{{\SetFigFont{9}{10.8}{\rmdefault}{\mddefault}{\updefault}$3$}}}}
\put(2200,-1861){\makebox(0,0)[rb]{\smash{{\SetFigFont{9}{10.8}{\rmdefault}{\mddefault}{\updefault}$1$}}}}
\put(2200,-3061){\makebox(0,0)[rb]{\smash{{\SetFigFont{9}{10.8}{\rmdefault}{\mddefault}{\updefault}$3$}}}}
\put(2200,-2461){\makebox(0,0)[rb]{\smash{{\SetFigFont{9}{10.8}{\rmdefault}{\mddefault}{\updefault}$2$}}}}
\put(2200,-3661){\makebox(0,0)[rb]{\smash{{\SetFigFont{9}{10.8}{\rmdefault}{\mddefault}{\updefault}$4$}}}}
\put(2333,-2086){\makebox(0,0)[b]{\smash{{\SetFigFont{9}{10.8}{\rmdefault}{\mddefault}{\updefault}$0$}}}}
\put(2333,-2686){\makebox(0,0)[b]{\smash{{\SetFigFont{9}{10.8}{\rmdefault}{\mddefault}{\updefault}$0$}}}}
\put(2333,-3286){\makebox(0,0)[b]{\smash{{\SetFigFont{9}{10.8}{\rmdefault}{\mddefault}{\updefault}$0$}}}}
\put(2333,-3886){\makebox(0,0)[b]{\smash{{\SetFigFont{9}{10.8}{\rmdefault}{\mddefault}{\updefault}$0$}}}}
\put(3796,-2086){\makebox(0,0)[b]{\smash{{\SetFigFont{9}{10.8}{\rmdefault}{\mddefault}{\updefault}$1$}}}}
\put(3663,-2686){\makebox(0,0)[b]{\smash{{\SetFigFont{9}{10.8}{\rmdefault}{\mddefault}{\updefault}$1$}}}}
\put(2732,-3286){\makebox(0,0)[b]{\smash{{\SetFigFont{9}{10.8}{\rmdefault}{\mddefault}{\updefault}$1$}}}}
\put(5659,-2086){\makebox(0,0)[b]{\smash{{\SetFigFont{9}{10.8}{\rmdefault}{\mddefault}{\updefault}$2$}}}}
\put(6058,-2686){\makebox(0,0)[b]{\smash{{\SetFigFont{9}{10.8}{\rmdefault}{\mddefault}{\updefault}$2$}}}}
\put(4195,-3886){\makebox(0,0)[b]{\smash{{\SetFigFont{9}{10.8}{\rmdefault}{\mddefault}{\updefault}$2$}}}}
\put(6856,-3286){\makebox(0,0)[b]{\smash{{\SetFigFont{9}{10.8}{\rmdefault}{\mddefault}{\updefault}$3$}}}}
\put(3397,-3886){\makebox(0,0)[b]{\smash{{\SetFigFont{9}{10.8}{\rmdefault}{\mddefault}{\updefault}$1$}}}}
\put(6058,-2986){\makebox(0,0)[lb]{\smash{{\SetFigFont{9}{10.8}{\rmdefault}{\mddefault}{\updefault}{\color[rgb]{0,0,0}$\xi$, $pulse_3(\xi) = 3$}%
}}}}
\end{picture}%

%% file: violatedsdc.pstex_t
\begin{picture}(0,0)%
\includegraphics{violatedsdc.pstex}%
\end{picture}%
\setlength{\unitlength}{3947sp}%
\begingroup\makeatletter\ifx\SetFigFont\undefined%
\gdef\SetFigFont#1#2#3#4#5{%
  \reset@font\fontsize{#1}{#2pt}%
  \fontfamily{#3}\fontseries{#4}\fontshape{#5}%
  \selectfont}%
\fi\endgroup%
\begin{picture}(3287,1231)(1526,-2735)
\put(2333,-2686){\makebox(0,0)[b]{\smash{{\SetFigFont{9}{10.8}{\rmdefault}{\mddefault}{\updefault}$\ell_i$}}}}
\put(1726,-1861){\makebox(0,0)[rb]{\smash{{\SetFigFont{9}{10.8}{\rmdefault}{\mddefault}{\updefault}$j$}}}}
\put(1726,-2461){\makebox(0,0)[rb]{\smash{{\SetFigFont{9}{10.8}{\rmdefault}{\mddefault}{\updefault}$i$}}}}
\put(1951,-1636){\makebox(0,0)[b]{\smash{{\SetFigFont{9}{10.8}{\rmdefault}{\mddefault}{\updefault}{\color[rgb]{0,0,0}$\xi_j$}%
}}}}
\put(2851,-2086){\makebox(0,0)[b]{\smash{{\SetFigFont{9}{10.8}{\rmdefault}{\mddefault}{\updefault}$\ell_j \leq \ell_i$}}}}
\end{picture}%

%% file: violatedspgc.pstex_t
\begin{picture}(0,0)%
\includegraphics{violatedspgc.pstex}%
\end{picture}%
\setlength{\unitlength}{3947sp}%
\begingroup\makeatletter\ifx\SetFigFont\undefined%
\gdef\SetFigFont#1#2#3#4#5{%
  \reset@font\fontsize{#1}{#2pt}%
  \fontfamily{#3}\fontseries{#4}\fontshape{#5}%
  \selectfont}%
\fi\endgroup%
\begin{picture}(4187,1156)(1526,-2735)
\put(2333,-2686){\makebox(0,0)[b]{\smash{{\SetFigFont{9}{10.8}{\rmdefault}{\mddefault}{\updefault}$\ell_i$}}}}
\put(1726,-1861){\makebox(0,0)[rb]{\smash{{\SetFigFont{9}{10.8}{\rmdefault}{\mddefault}{\updefault}$j$}}}}
\put(1726,-2461){\makebox(0,0)[rb]{\smash{{\SetFigFont{9}{10.8}{\rmdefault}{\mddefault}{\updefault}$i$}}}}
\put(4951,-1711){\makebox(0,0)[b]{\smash{{\SetFigFont{9}{10.8}{\rmdefault}{\mddefault}{\updefault}{\color[rgb]{0,0,0}$\xi_j$}%
}}}}
\put(3226,-2086){\makebox(0,0)[b]{\smash{{\SetFigFont{9}{10.8}{\rmdefault}{\mddefault}{\updefault}$\ell_j = \ell_i + 1$}}}}
\put(3751,-1711){\makebox(0,0)[lb]{\smash{{\SetFigFont{9}{10.8}{\rmdefault}{\mddefault}{\updefault}{\color[rgb]{0,0,0}$S_j(\ell_j)$}%
}}}}
\end{picture}%

%% file: conc.tex
\section{Concluding remarks}
\label{sec:conc}

Partially ordering the executions of a distributed algorithm is a mechanism to restrict its set of executions. In several cases, this set of restricted executions comprises more efficient executions than its complement. We have presented partially ordered executions of a distributed algorithm as the executions satisfying some restricted orders of their actions in two different frameworks, those of event- and pulse-driven computations. In the event-driven framework, we have given new conditions for message delivery that generalize the ones leading to FIFO ordering and to causal ordering. An important property of these generalized conditions is that they can be dynamically tuned to become more or less strict as the computation evolves. The same principle has been applied to the pulse-driven framework, in which case a constraint on the relation between the pulse generation mechanism and message delivery is established to generalize the well-known fully synchronous ordering.

The algorithm which partially orders the executions in each case may introduce some overhead, which is large to the same extent that the order restrictions are strict (the extremal case is that of causal ordering, in the event-driven framework, or that of fully synchronous ordering, in the pulse-driven case). Efficient implementations are application-dependent and correspond to those implementations that lead to computations likely to be efficient with limited overhead. For instance, an efficient implementation of the partially ordered version of randomized distributed backtrack search will provide a satisfactory trade-off between the number of messages postponed and the number of unsuccessful donation requests. In the same vein, an implementation of the distributed iterative algorithm for systems of linear equations will be efficient when the gain in the number of iterations surpasses the overhead of ordering the execution. We expect that systematic experimentation on real-world instances of both problems will yield crucial insight into the most appropriate choices.